\documentclass{PoS}
\usepackage{amsmath,bm}
\usepackage{epsfig}
\usepackage{amssymb}
\usepackage{fontenc}
\usepackage{mathptmx}
\usepackage{graphicx}

\def\be{\begin{equation}}
\def\ee{\end{equation}}
\def\bea{\begin{eqnarray}}
\def\eea{\end{eqnarray}}

\usepackage{graphicx}
\def\lesssim{\ \hbox{\raise 2pt \hbox{$<$} \kern -13pt
                     \lower 3pt \hbox{$\sim$}}\ }
\def\greatersim{\ \hbox{\raise 2pt \hbox{$>$} \kern -13pt
                     \lower 3pt \hbox{$\sim$}}\ }

\def\herwig{{\sc Herwig}}
\def\pythia{{\sc Pythia}}

\def\desepsf(#1 width #2){\epsfxsize=#2 \epsfbox{#1}}

\title{Studies of forward jets and production of W, Z bosons at LHC energies}

\ShortTitle{Studies of forward jets and production of W,Z}

\author{K. Kutak \\
       Deutsches Elektronen Synchrotron, D-22603 Hamburg, Germany\\
       E-mail: \email{kutak@mail.desy.de}}

\abstract{We report on application of QCD in calculations of forward  jet  
and W, Z boson production  cross sections at the Large Hadron Collider.  In particular 
in case of jet production we emphasize  dynamical features  
of   the  matrix elements   controlling   the resummation   
of  logarithmically enhanced  corrections  in  $\sqrt{s}/E_T$, where $E_T$ is the jet production transverse 
energy. In case of production of W, Z bosons we focus on angular correlations between produced boson and hardest
associated jet.}

\FullConference{The 2009 Europhysics Conference on High Energy Physics,\\
		 July 16 - 22 2009\\
		 Krakow, Poland}

\begin{document}
\section{Introduction}
Experiments at the Large Hadron Collider  (LHC) will allow to test standard model at very high energies. Here we are interested in Quantum Chromodynamics (QCD)
processes like forward jet production and interplay of QCD with Electro-Weak interactions which lead to $W$, $Z$ bosons production in hadronic collision. 
The first of listed processes is of interest since it will allow for better understanding of partonic structure of the proton at extreme energies 
where new phenomena like saturation~\cite{sat, sat2, dent} may occur. The other process is an important candle of Electro-Weak theory which has to be tested at 
various energy scales in order to have full understanding of it. This process will also serve as one to calibrate calorimeters and to 
measure luminosities.\\
The large center of mass energy at the LHC will require a need to apply QCD resummation approaches capable to account for multiple scales 
in the problem. Namely, one has to account for logarithms of type $\alpha_s\ln(\mu/\Lambda_{QCD})$ where $\mu$ might be $p_T$ of jet or mass of $W/Z$  and 
logarithms coming from the fact that at least parton densities in one of the incoming proton will be probed at very small longitudinal momentum fraction 
$x$ giving rise to logarithms of type $\alpha_s^n\ln^m1/x$. The theoretical framework to resume consistently both kinds of logarithmic corrections  in QCD calculations   
is based on  high-energy factorization at fixed transverse momentum.~\cite{hef}. 
This formulation    depends  on  unintegrated distributions for 
parton splitting, obeying appropriate evolution equations, and short-distance, 
process-dependent matrix elements.
The  unintegrated-level evolution is given by  evolution equations in rapidity,   or angle,  parameters.   
Different  forms of the  evolution, valid  in   different kinematic regions, are available, see~\cite{jcc-lc08,fhfeb07}, 
and references therein, for recent work  in this area and reviews. 
In this article we present recently obtained results  for hard matrix elements needed in application of factorisation formulas 
for both of introduced above processes. 
In Sec.~2 we introduce  the basic structure of  jet production in the LHC forward region. In Sec.~2.1 we consider   associated 
parton showering effects.  In Sec.~2.3  we consider effects from  the short-distance matrix elements that control the resummation 
of logarithmically enhanced corrections in $\sqrt{s} / E_T$, where $E_T$ is the hard jet transverse energy. In Sec. 3 we briefly discuss central production of
$W$, $Z$ bosons focusing on observables which are sensitive to different showering methods and different assumptions on initial state of colliding partons.  
We give concluding remarks in Sec.~4.

\section{Forward jets}
\begin{figure}[htb]
\vspace{55mm}
\includegraphics{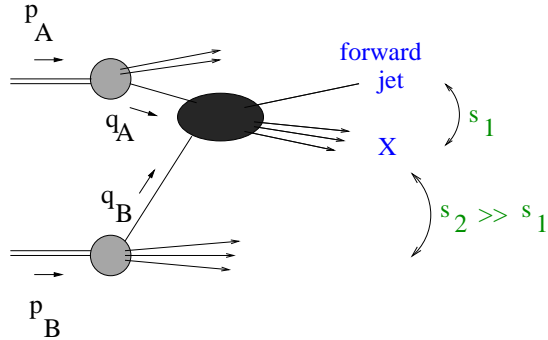}
\caption{Jet  production in the forward rapidity region 
in  hadron-hadron collisions.} 
\label{fig:forwp}
\end{figure}

\begin{figure}[htb]
\vspace{30mm}
\includegraphics{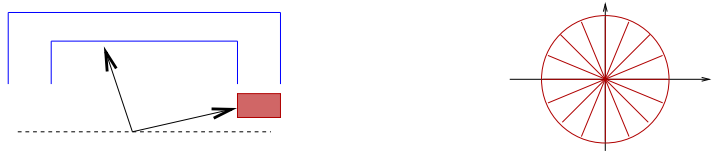}
\caption{(Left)  High-$p_\perp$  events  in the 
forward and central detectors; (right) azimuthal plane segmentation.} 
\label{fig:azimcorr}
\end{figure}

 The hadroproduction of a  forward jet   associated  
with  hard final state $X$  is pictured 
in Fig.~\ref{fig:forwp}.    
The kinematics 
  of the process     is  characterized 
by the  large  ratio  of sub-energies  $s_2  /  s_1 \gg 1 $   
 and  highly asymmetric longitudinal momenta in the partonic initial 
  state, $q_A \cdot p_B \gg q_B \cdot p_A$.  
 At the LHC the use of  forward calorimeters  allows  one to  
  measure    events  where   jet transverse momenta 
  $p_\perp  >   20$ GeV   are produced  several units of rapidity 
  apart,  $\Delta y   \greatersim 4 \div 6$~\cite{cmsfwd,aslano,heralhc}.  
 Working at    polar angles that are  small   but   sufficiently  far  from the beam axis 
 not to be affected by  beam remnants,     one measures 
 azimuthal plane correlations  
  between   high-$p_\perp$  events   (Fig.~\ref{fig:azimcorr})
  widely  separated    in rapidity~\cite{heralhc,preprint}.

The   presence of multiple   large-momentum scales  
implies   that, as      recognized in~\cite{muenav,vddtang,stirl94},    
reliable  theoretical predictions   for forward jets 
can only be obtained after  summing  
 logarithmic  QCD corrections at high energy
 to all orders in $\alpha_s$\footnote{Analogous observation applies to  
  forward jets  associated 
 to  deeply    inelastic  scattering~\cite{mueproc90c,forwdis92}.    Indeed, measurements of 
 forward jet cross sections at Hera~\cite{heraforw} have illustrated that 
 either fixed-order next-to-leading  
 calculations  or  standard shower 
 Monte Carlos~\cite{heraforw,web95,webetal99}, e.g.   \pythia\ 
  or \herwig,  are not  able to   describe 
 forward jet  $ep$ data.}.         
This    motivates    efforts~\cite{webetal99,orrsti,stirvdd,andsab}  to 
 construct   new,   improved  
  algorithms  for   Monte Carlo  event  generators capable of 
   describing   jet   production    beyond  the central rapidity region.

 In the  LHC forward   kinematics,  
  realistic  phenomenology  of      hadronic  jet final states 
  requires taking   account    of      
 both logarithms  of the large  rapidity  interval  
(of  high-energy type)  
and logarithms of the hard transverse momentum (of collinear type).  
The theoretical framework to resume   consistently 
both  kinds of logarithmic corrections  in QCD 
 calculations   is based on  high-energy    factorization at 
fixed transverse momentum~\cite{hef}.    

 Ref.~\cite{preprint}     
  investigates  forward jets  in this framework. It presents  the 
  short-distance matrix elements   needed to evaluate the 
   factorization formula, including  all partonic channels,  
   in a fully exclusive form.   
   On one hand, once convoluted with the BFKL off-shell gluon Green's function 
 according to the method 
 of~\cite{hef}, these matrix elements 
  control the summation of high-energy 
 logarithmic corrections to the jet cross sections. They 
 contain   contributions   both 
 to the   next-to-leading-order BFKL kernel~\cite{fadlip98} 
 and to the jet impact factors~\cite{mc98,schw0703}.  
   On  the other hand, they can   be used in a shower 
   Monte Carlo  generator    implementing  parton-branching kernels 
   at unintegrated level (see e.g.~\cite{jadach09,hj_ang}  for recent works)  
   to   generate fully exclusive events.

\begin{figure}[htb]
\vspace{45mm}
\includegraphics{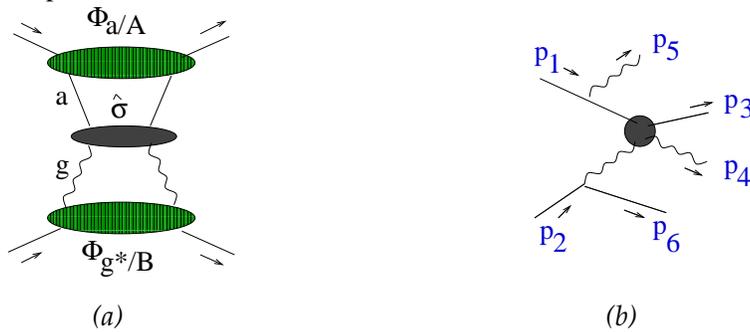}
\caption{(a) Factorized structure of the cross section; (b) a typical contribution  
 to the 
$q g$ channel matrix element.} 
\label{fig:sec2}
\end{figure}

The   high-energy  factorized 
form~\cite{hef,preprint,mc98}      
 of the  forward-jet  cross section is   represented   in Fig.~\ref{fig:sec2}a. 
Initial-state parton configurations  contributing to  
forward production are asymmetric, 
with the parton in the top subgraph being  probed near  the mass shell and  
large  $ x $,  
while  the parton in  the bottom subgraph is off-shell and small-$x$. 
The    jet  cross  section differential 
in the final-state   
transverse  momentum 
 $Q_\perp$  and  azimuthal angle $\varphi$ 
is given  schematically  by~\cite{hef,preprint,mc98}  
\begin{equation}
\label{forwsigma}
   {{d   \sigma  } \over 
{ d Q_\perp^2 d \varphi}} =  \sum_a  \int  \    \phi_{a/A}  \  \otimes \  
 {{d   {\widehat  \sigma}   } \over 
{ d Q_\perp^2 d \varphi  }}    \  \otimes \   
\phi_{g^*/B}    \;\; , 
\end{equation}
where  
$\otimes$ specifies  a convolution in both longitudinal and transverse momenta, 
$ {\widehat  \sigma} $  is the  hard scattering  cross section,  calculable 
 from  a  suitable off-shell continuation of 
perturbative   matrix elements,  $ \phi_{a/A} $ is the distribution of parton 
$a$ in hadron $A$ 
obtained from   near-collinear shower evolution, and $ \phi_{g^*/B} $ is  
  the gluon unintegrated distribution in hadron $B$ 
  obtained from non-collinear, 
  transverse momentum  dependent shower evolution. 
 
In the  next section we  comment on the initial-state shower evolution. 
In Sec.~2.2 we turn to   hard-scattering  contributions.

\subsection{Parton shower evolution} 

\begin{figure}[htb]
\vspace{45mm}
\includegraphics{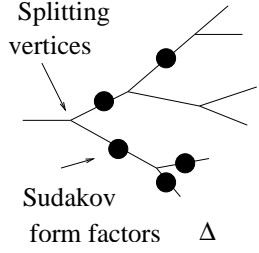}
\caption{Parton  branching in terms of splitting probabilities and form factors.}
\label{fig:pshower}
\end{figure}

Parton distributions can be obtained  
by parton-shower Monte Carlo methods   
  via   branching algorithms 
based  on collinear evolution 
of the jets developing    
from  the hard event~\cite{mc_lectures}. The branching probability  can be   given 
 in terms    of  two basic 
quantities  (Fig.~\ref{fig:pshower}), the  splitting functions   
 at the vertices of the parton cascade     and the  form 
 factors   to go from one vertex  to the other. 
An important ingredient of this approach is the inclusion of  soft-gluon 
coherence  effects~\cite{mc_lectures,dokrev,mc89} through angular ordering of the 
emissions in the shower.   

\begin{figure}[htb]
\vspace{36mm}
\includegraphics{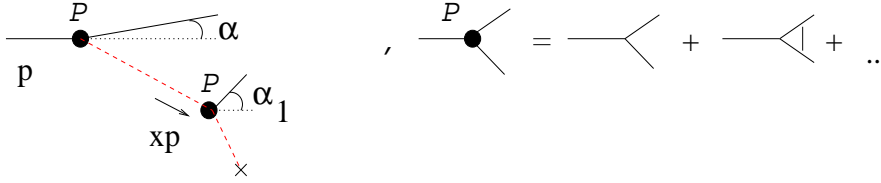}
\caption{(left) Coherent radiation 
 in the space-like parton shower for $x \ll 1$; (right) the unintegrated 
splitting function ${\cal P}$, including small-$x$ virtual 
corrections.} 
\label{fig:coh}
\end{figure}

Corrections to collinear-ordered showers, however,  arise 
 in  high-energy processes with multiple  
 hard scales~\cite{heralhc,mw92,bo04},  as is the case  with 
 the  production  of jets at  forward rapidities in Fig.~\ref{fig:forwp}.
 In particular, new  color-coherence   effects  set in  in this regime 
 due to    emissions  from    internal lines in the  
 branching decay chain~\cite{heralhc,mc98,anderss96}  
    that  involve
 space-like partons   
 carrying  small   longitudinal momentum   
 fractions.   
 The picture of the coherent branching   is modified   in this case because 
 the emission  currents become dependent on the total transverse 
 momentum  transmitted down the  initial-state parton decay 
   chain~\cite{hef,mc98,mw92,bo04,jung04}.  Correspondingly,   
   one needs to work 
   at the level of    unintegrated   
   splitting functions and  partonic distributions~\cite{jcc-lc08,hj_rec}  
   in order to take into account 
     color coherence not only  for  large 
 $x $ but also  for  small $x$  in the angular region (Fig.~\ref{fig:coh}) 
\begin{equation}
\label{cohregion}
\alpha / x  > \alpha_1 > \alpha \hspace*{0.3 cm}    , 
\end{equation} 
where the angles  $\alpha$ for the 
partons radiated from the initial-state 
shower are taken with respect to the 
initial beam jet direction, and increase with increasing off-shellness.

The case of  LHC forward  jet production is a multiple-scale problem where coherence effects  of the kind  above  
enter, in the factorization formula  (\ref{forwsigma}), 
 both the  short-distance factor $ {\widehat  \sigma} $ and   the 
long-distance factor  $\phi$.
 Contributions   from  the coherence   region   
(\ref{cohregion})  are potentially enhanced by   terms 
$ 
\alpha_s^n \ln^{m}  \sqrt{s} / p_\perp   $ 
where $\sqrt{s}$ is the total center-of-mass energy and $p_\perp$  is  
the   jet  transverse momentum\footnote{Terms with 
 $m  > n$   are known to 
   drop out  from inclusive  processes due to strong cancellations 
associated with coherence,    
   so that,  for instance,  the 
    anomalous dimensions $\gamma^{ i j}$  for   
 space-like evolution    receive   at most   single-logarithmic 
 corrections at high energy~\cite{fadlip98,ch94}. This need not be the case for  
 exclusive  jet  distributions, where   such cancellations  are not  present 
and one may expect   larger enhancements.}.  
These contributions represent corrections to 
the  angular  ordering  implemented in  collinear  showers and 
are not included   at present 
in standard Monte Carlo generators~\cite{mc_lectures}. 
 Work  to develop 
 methods for   unintegrated shower evolution, capable of including 
 such corrections,  is 
   underway  by  several authors. 

The  proposal~\cite{jadach09} incorporates 
NLO  corrections to  flavor non-singlet  QCD evolution  in 
  an unintegrated-level Monte Carlo. 
The approach is based on the generalized ladder expansion of~\cite{CFP}, which is 
   extended to the high-energy   
region  in~\cite{ch94}. This approach   could  in principle   be applied generally,  
including  flavor  singlet evolution, and used 
to treat also forward hard processes.

\subsection{The factorizing hard cross sections}

 Logarithmic corrections 
 for large rapidity $y \sim  \ln s / p_\perp^2$ 
are  resummed to all orders in $\alpha_s$ 
 via Eq.~(\ref{forwsigma}),    
   by convoluting     (Fig.~\ref{fig:sec2})  unintegrated distribution 
 functions  with   
 well-prescribed short-distance matrix elements, 
 obtained from the high-energy limit of  higher-order 
 scattering amplitudes~\cite{preprint,mc98}.  
With reference to     Fig.~\ref{fig:sec2}b, 
in the   forward production region we have 
  $(p_4+ p_6)^2  \gg  (p_3 +p_4)^2   $     and 
 longitudinal momentum ordering,  so that 
\begin{equation}
\label{fwdkin}
p_5 \simeq  (1 - \xi_1 ) p_1 \;\;\;   ,  \;\;\;\;\;  p_6 \simeq   (1 - \xi_2 ) p_2 - k_\perp   
 \;\;\;   ,  \;\;\;\;\;  
\xi_1 \gg \xi_2     \;\;  .  
\end{equation}
Here $\xi_1$ and $\xi_2$ are longitudinal momentum fractions,   and $ k_\perp $  is the 
di-jet  transverse momentum in the laboratory frame.   
It is convenient to  define  the rapidity-weighted average
$Q_\perp = (1-\nu) p_{\perp  4} - \nu p_{\perp  3}$, with  
$\nu =  (p_2 \cdot p_4) / p_2 \cdot (p_1 -p_5) $.  In    Fig.~\ref{fig:sec2}b  
Eq.~(\ref{forwsigma})  factorizes  the high-energy $q g$ amplitude 
 in front       of the (unintegrated) distribution from 
 the splitting in the bottom subgraph. 
The factorization in terms of   this    parton 
splitting    distribution  is valid at large $y$      not only in  the collinear  region 
but also in the large-angle emission region~\cite{hef}.    As a result  
  the rapidity resummation is  carried out  consistently with 
perturbative  high-$Q_\perp$ corrections~\cite{hef,mc98} at any fixed 
order in $\alpha_s$.

\begin{figure}[htb]
\vspace{44mm}
\includegraphics{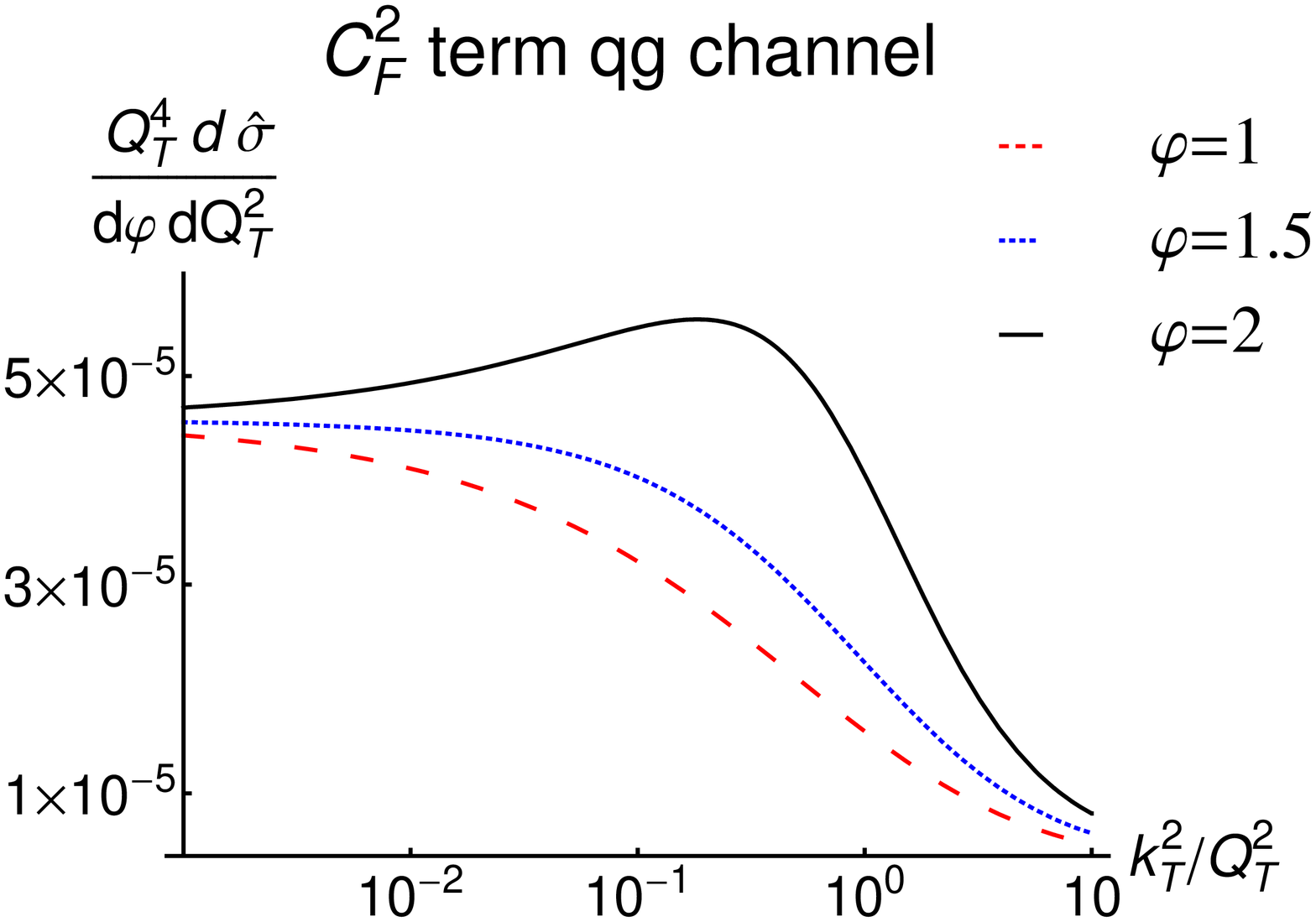}
\includegraphics{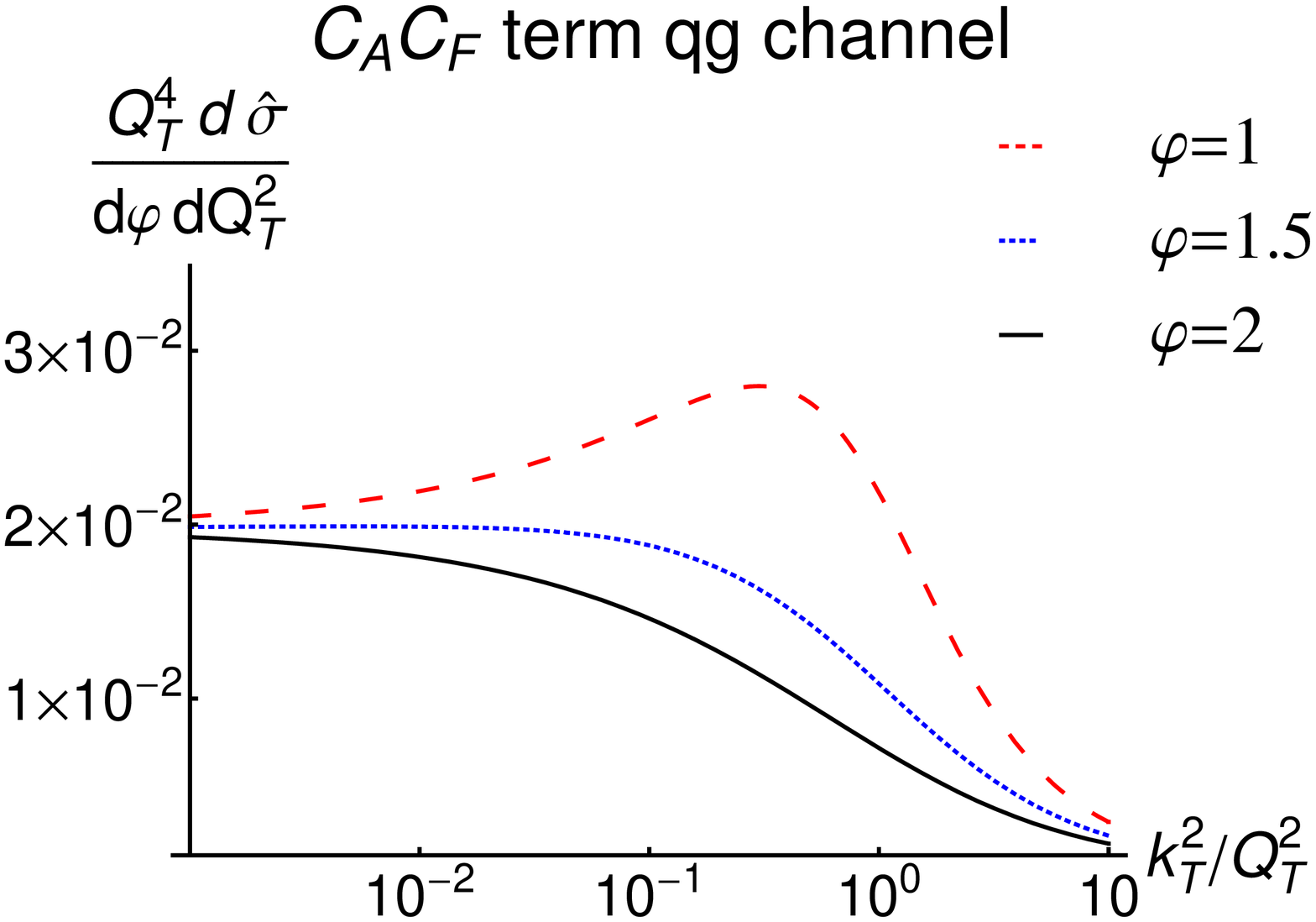}
\caption{The  $k_T / Q_T$  dependence of the  
factorizing   $q g$   hard  cross 
section at high energy~\cite{preprint}:           
 (left) $C_F^2$ term;  (right) $C_F   
C_A$ term.} 
\label{fig:forwplot}
\end{figure}

\begin{figure}[htb]
\vspace{44mm}
\includegraphics{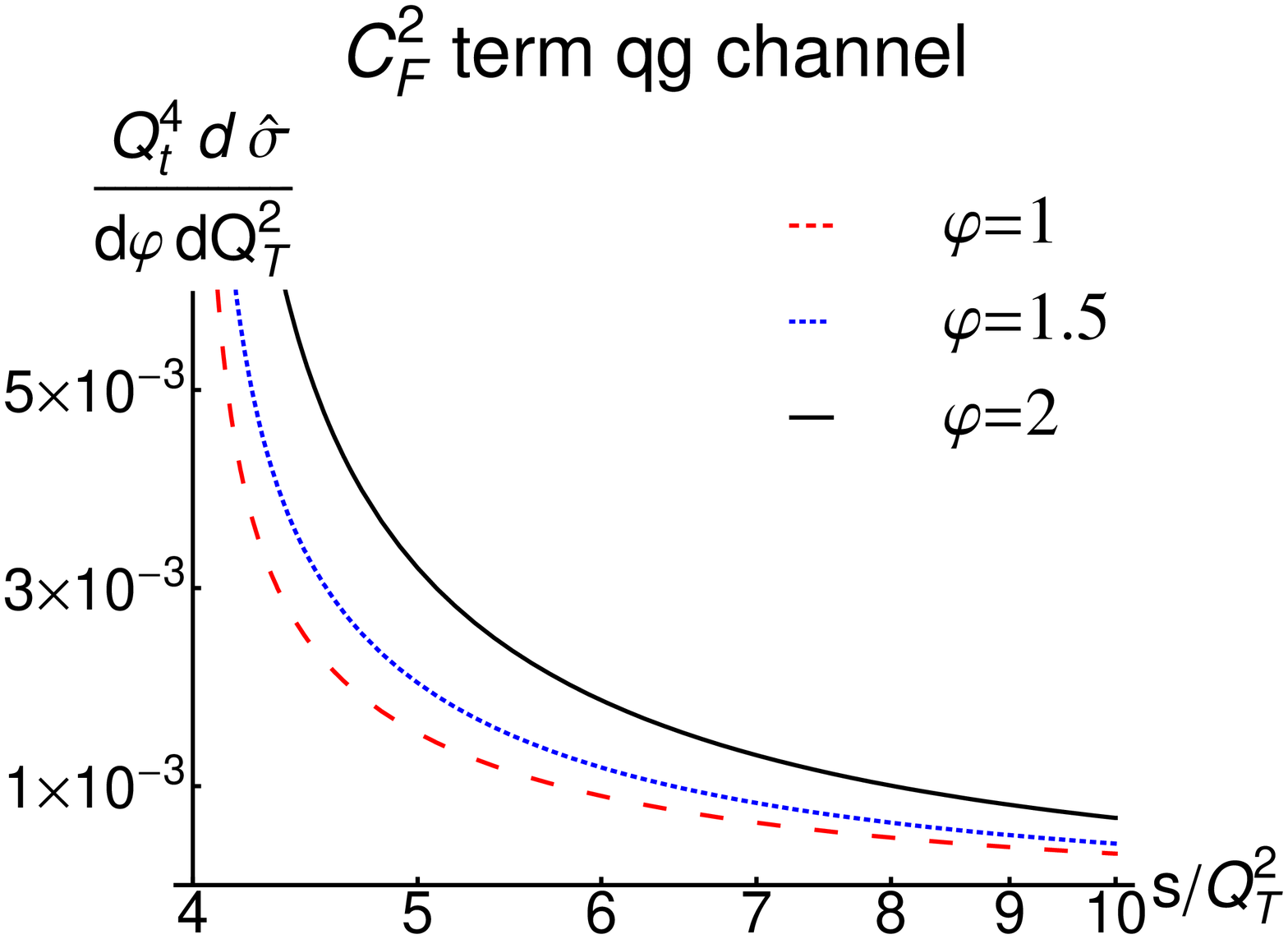}
\includegraphics{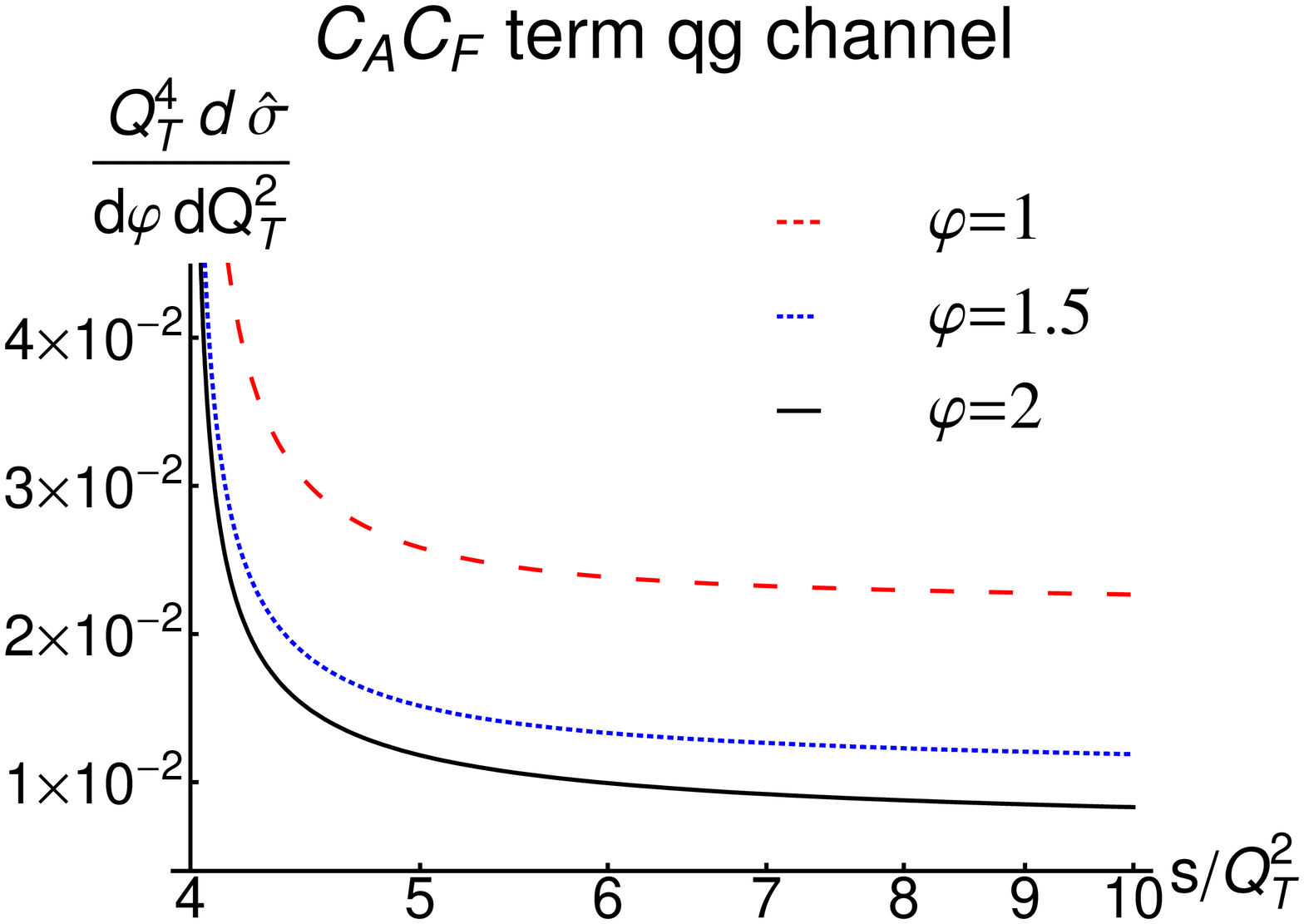}
\caption{The  energy dependence of the    $q g$ hard cross section~\cite{preprint} 
($  k_T / Q_T =1$).} 
\label{fig:forwplot1}
\end{figure}

The explicit expressions  for  the relevant 
high-energy  amplitudes  
are  given in~\cite{preprint}. 
Figs.~\ref{fig:forwplot}  and \ref{fig:forwplot1}  illustrate  features of the 
factorizing matrix elements, partially integrated over final states. We  plot  
distributions differential in  $Q_\perp$ and  azimuthal angle $\varphi$ 
($\cos \varphi = Q_\perp \cdot k_\perp / |  Q_\perp |  | k_\perp | $) for the case 
of the $q g$ channel.   
Fig.~\ref{fig:forwplot} shows the dependence on  $k_\perp$,  which measures 
  the distribution  
  of   the  third    jet   recoiling against the leading di-jet system. 
    Fig.~\ref{fig:forwplot1}   shows the energy dependence.

The region    $ k_\perp / Q_\perp   \to  0$  in Fig.~\ref{fig:forwplot}  
corresponds to 
the leading-order   process with two back-to-back 
jets.   
The resummation  of  the higher-order  logarithmic corrections 
 for large  $y \sim  \ln s / p_\perp^2$  is precisely  
 determined~\cite{hef,mc98}    by  integrating the 
 u-pdfs   over the  $ k_\perp$-distribution  in  Fig.~\ref{fig:forwplot}.   
So the results       in   
Fig.~\ref{fig:forwplot}  
 illustrate    quantitatively  the  significance of  
contributions  
with   $k_\perp \simeq  Q_\perp$ in the large-$y$   region.    
  The role of coherence  from multi-gluon emission 
  is to set the  dynamical cut-off     at values of 
  $ k_\perp $ of order  $ Q_\perp $. 
  Non-negligible 
  effects arise at high energy    from the finite-$k_\perp $  tail. 
These  effects are not included in  collinear-branching  
 generators  (and only partially in fixed-order 
 perturbative calculations),   
 and  become  more and more important  as  the jets are  observed at 
large rapidity separations.   The   dependence on the azimuthal angle 
in    Figs.~\ref{fig:forwplot}  and \ref{fig:forwplot1} 
is  also relevant, as   forward  jet    measurements  will rely 
  on azimuthal plane correlations between  jets  
far apart  in rapidity (Fig.\ref{fig:azimcorr}). 
  
  Results for  all  other partonic channels   are given    in~\cite{preprint}.   
After including parton showering~\cite{prepar},    quark- 
  and gluon-initiated  contributions   are of comparable size 
 in  the 
  LHC forward kinematics:  
    realistic phenomenology requires including all channels. 
  Note  also that 
since the forward kinematics selects 
asymmetric parton momentum fractions,   effects   
     due to    the   $ x \to 1$   endpoint   behavior~\cite{fhfeb07}   
  at  the 
fully  unintegrated level  may become relevant as well.  

\section{W, Z boson production}  
Central production of $W/Z$ bosons at LHC energies will be dominated by gluonic component of partonic system.
The longitudinal components of incoming gluons four momenta will be small and of the same order what will result in 
the central production.
As in the forward jet case the framework to consistently account for this kinematical set up is provided by $k_\perp$ factorisation approach
where 
$\sigma(g^*g^*\to q(W/Z)\overline{q})$ is an off shell continuation of hard matrix element for $\sigma(gg\to q(W/Z)\overline{q})$ \cite{deak,zotov} 
which allows to include effects coming from finite transversal momenta of gluons.
The interesting observable to calculate for this production process is the angular distance between highest $p_\perp$ jet and $p_\perp$ of $Z$ or $W$,
which allows for estimation of uncertainties of theoretical predictions. On Fig. \ref{fig:ptz} (left) we show comparison
of calculation based on $k_\perp$ factorisation approach \cite{deak,zotov} and collinear approach at LO and NLO. We see that the distributions are 
considerably different for
small angles in case of $k_\perp$ factorisation approach and collinear LO. While NLO collinear is quite similar to the one obtained in $k_\perp$ factorisation
approach.
The reason for that comes from the momentum conservation in LO collinear approach which forbids events in region from $0$ to $\pi/2$. This not the case in $k_\perp$
factorisation approach and NLO collinear calculation where the additional transversal momentum flow allows momenta of $Z$, $b$ and $\overline b$ to be
unbalanced.  
\begin{figure}[t!]
\vspace{55mm}
\includegraphics{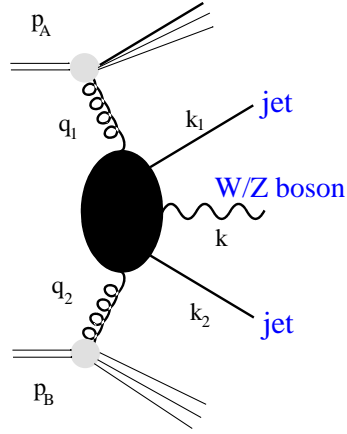}
\caption{Jet  production in the forward rapidity region 
in  hadron-hadron collisions.} 
\label{fig:forwpicture}
\end{figure}
The $k_\perp$ factorisation formula while applied to this process takes form:
\begin{equation}
\label{bosons}
\sigma=\phi_{a^*/A}  \  \otimes \widehat  \sigma(g^*g^*\to q(W/Z)\,\overline{q}) \phi_{b^*/B}    \;\; ,
\end{equation}
The other important distribution is the differential cross section in the transversal momentum of $Z$ $b\overline b$ system \ref{fig:ptz} (right). This distributions 
are rather different at small $p_{Zb\overline b\perp}$ region. The difference is due to the fact that in the $k_\perp$ factorisation approach subleading terms 
of all orders from point of view of collinear approach are taken into account. We also see that at higher values of $p_{Zb\overline b\perp}$ both approaches agree.   
\begin{figure}[t!]
{\includegraphics[width=0.49\columnwidth]{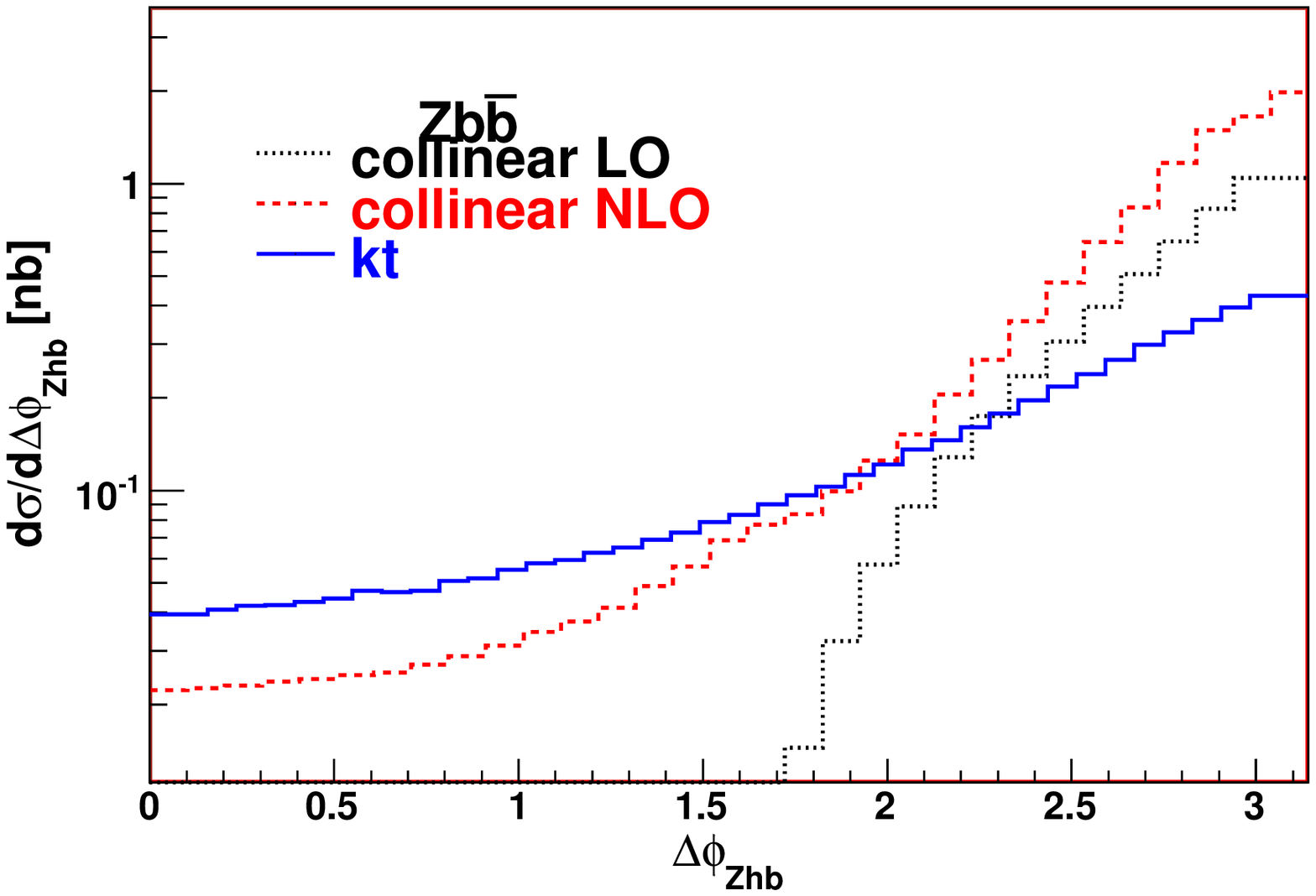}}
{\includegraphics[width=0.49\columnwidth]{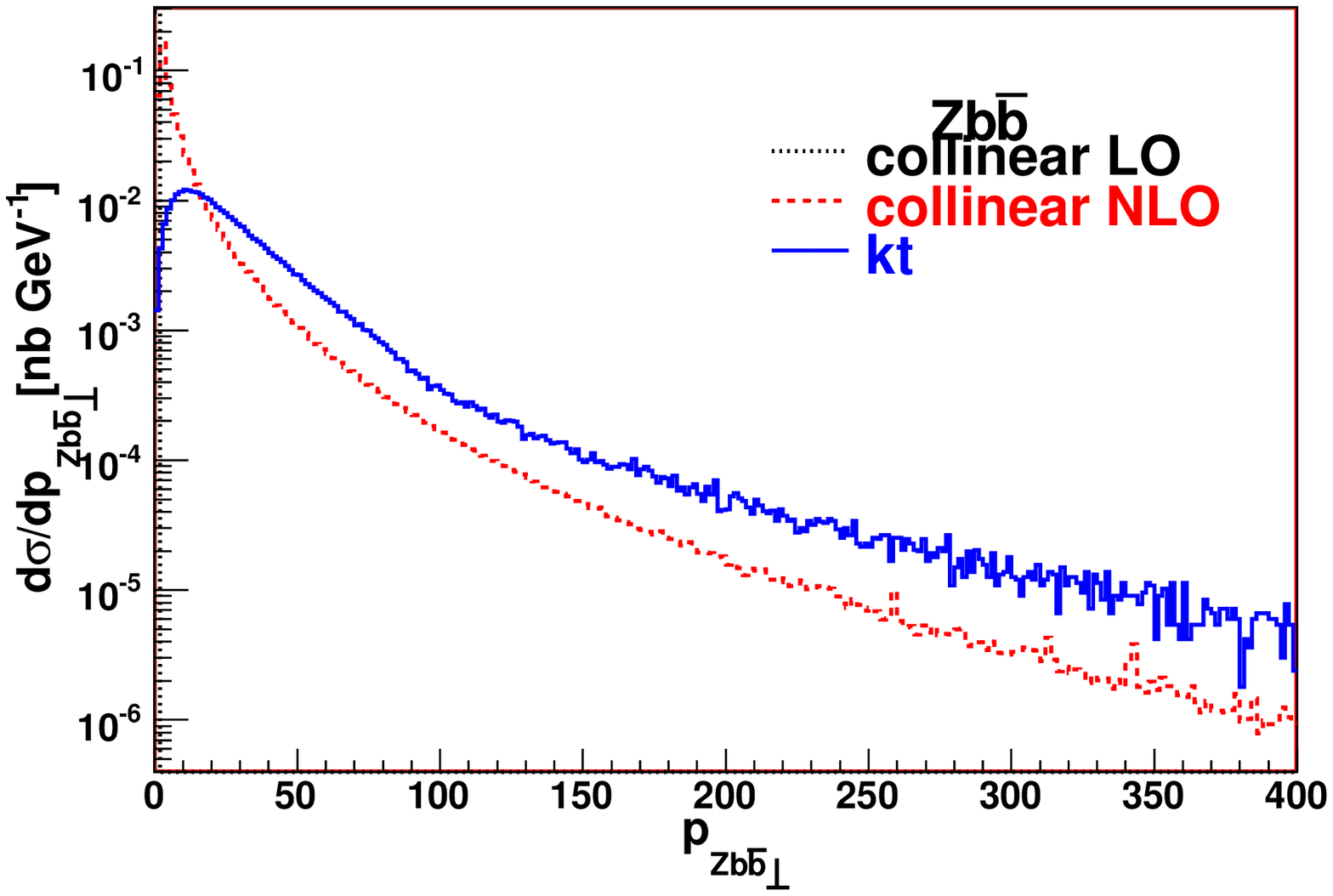}}
\caption{Distributions of the distance in azimuthal angle of Z and highest p$_\perp$ quark or antiquark. Calculation with massive $b$-quarks (left). 
Comparison of cross sections in transverse momentum of the produced $Z$ gauge boson. Calculation with massless $b$-quarks (right)}\label{fig:ptz}
\vspace*{-0.2cm}
\label{fig:ptz}
\end{figure}
\section{Conclusions}
Forward + central detectors  at the LHC  allow  jet   correlations  to be  measured across  rapidity intervals of  several  units,   
 $\Delta y   \greatersim 4 \div 6$.   Such multi-jet    states   can   be relevant to new particle  discovery processes as well as 
 new aspects   of  standard model physics.  
 Existing sets of  forward-jet  data in ep collisions, much more limited  than the potential LHC  yield, indicate  that neither 
 conventional parton-showering Monte Carlos nor next-to-leading-order   QCD calculations  are capable of  describing forward jet    phenomenology.  
Improved methods to evaluate QCD    predictions are needed to treat the multi-scale region implied by  the  forward kinematics.   
In this article we have discussed ongoing progress,  examining in particular   factorization properties of multi-parton matrix elements in the forward region, 
and prospects  to include  parton-showering effects with gluon coherence  not only in the collinear region but also   in the large-angle emission region.
We also have discussed predictions for $W$ and $Z$ central production based on different showering schemes. 
\newpage
\acknowledgments  
 I thank the  conference organizers   and    the conference staff   
for the nice atmosphere at the meeting. 
The results presented in Sec. 2 of this article have been obtained   
in collaboration with M.~De\'ak, F.~Hautmann and H.~Jung while the results presented in Sec.3 were obtained by M. De\'ak and F. Schwennsen

\end{document}